\documentclass[aps, pre, twocolumn, showpacs, nobibnotes, 10pt]{revtex4-1}

\catcode`,\active

\catcode`\,12

\usepackage{graphicx, amsmath, amssymb, bm, mathtools}

\newcommand{\be}{\begin{equation}}
\newcommand{\ee}{\end{equation}}
\newcommand{\BE}{\begin{eqnarray}}
\newcommand{\EE}{\end{eqnarray}}

\usepackage{ifpdf}
\ifpdf
    \newcommand{\ignoreThis}[1]{}
    \usepackage{color}
    \definecolor{Gray}{rgb}{0.6,0,0}
\else
    \newcommand{\ignoreThis}[1]{#1}
    \usepackage[usenames,dvips]{color}
\fi
\usepackage[normalem]{ulem}

\begin{document}

\title{A framework for analyzing ecological trait-based models in multi-dimensional niche spaces}

\author{Tommaso Biancalani$^1$}
\author{Lee DeVille$^2$}
\author{Nigel Goldenfeld$^1$}

\affiliation{$^1$Department of Physics and Institute for Genomic
Biology, University of Illinois at Urbana-Champaign, Loomis Laboratory
of Physics, 1110 West Green Street, Urbana, Illinois 61801-3080, USA}

\affiliation{$^2$Department of Mathematics and Institute for Genomic
Biology, 1409 West Green Street, Urbana, IL 61801}

\begin{abstract}
We develop a theoretical framework for analyzing ecological models with a multi-dimensional niche space. The novelty of our approach relies on the fact that ecological niches are described by sequences of symbols, which allows us to include multiple phenotypic traits. Ecological drivers, such as competitive exclusion, are modelled by introducing the Hamming distance between two sequences. We show that a suitable transform diagonalizes the community interaction matrix of these models, making it possible to predict the conditions for niche differentiation and the asymptotically long time population distributions of niches. We exemplify our method using the Lotka-Volterra equations and suggest a procedure to establish contact with experimental data.

\end{abstract}


\maketitle

\section{Introduction}

Since before ``ecology'' had a name it had been recognized that the
manner in which species occupying a given ecosystem exploit that
system's resources is distributed in a markedly structured, non-uniform
manner.  More-or-less distinct ecological niches are occupied to the
exclusion of apparently available intermediate
strategies~\cite{odum2005fundamentals}. This observation has led to a
long history of observational and modeling studies in the effort to
understand, and hopefully be able to predict, the structure,
complexity, and stability of niche
occupation~\cite{odum2005fundamentals}.

Although it is clear that resource competition is an essential driver
of ecological structure~\cite{hardin1960competitive}, thus leading to
niche differentiation and specialization, it is equally clear that this
idea by itself cannot be the whole story. In particular it leaves many
unanswered questions about the detailed structure, reproducibility, and
dynamic properties of niche occupation. Are there, for example, limits
to the number of different niches a given environment can
support?~\cite{macarthur1967limiting}  Are niches stochastic and
emergent phenomena \cite{odling2003niche}, are they environmentally
dictated, or is a variable interplay of both factors
involved?~\cite{grant2002unpredictable}

A particularly interesting set of such questions arise from the common
observation that very closely related species often co-exist in the
same environment apparently occupying very nearly the same, if not
identical, niches. It is clear from field studies and theoretical
investigations that multiple factors are involved in this aspect of
niche evolution. For example, the niche space is highly dimensional,
allowing the individuals to minimize competition by moving into the
various directions of the niche space. A classic example is that of the
various types of \textit{Anolis} lizards found in tropical rainforests,
which share a common prey -- insects -- but avoid competition by living
in different parts of the rainforest~\cite{roughgarden1974niche}.
Various species of finches look similar to each other except for such
traits as beak design, which have specialized the finches to different
food needs. Analogous observations have been made in (e.g.)
plants~\cite{shmida1985coexistence} and
animals~\cite{strong1982harmonious} and soil microbial
communities~\cite{lennon2012mapping}.

Another, less intuitive, phenomenon, is that similar species can also
coexist in a one-dimensional niche space, as they suppress shared
competitors~\cite{scheffer2006self}. This latter claim has been
supported by Lotka--Volterra models exhibiting ``lumped distributions''
(sometimes referred to as \lq\lq clumps"), i.e. clusters of similar
individuals separated by exclusion zones. Their emergence has been
analyzed in further mathematical studies~\cite{pigolotti2007species,
hernandez2009species, pigolotti2010gaussian} which have shown that, in
these models, lumped distributions arise due to an underlying pattern
instability.  This instability in niche space arises from competitive
exclusion, and can be thought of as individuals between clumps
experiencing double competition from each of the neighboring clumps,
thus suppressing their growth relative to the clumps. The outcome of
these interactions will of course depend on the type of competition
kernel employed in the Lotka-Volterra model and has been parameterized
and analyzed in previous works~\cite{scheffer2006self,pigolotti2007species,
hernandez2009species, pigolotti2010gaussian}.

Until now, lumped distributions been investigated in a one-dimensional
niche space, but not in multidimensional counterparts. This could
potentially be important, because the domain of competition now becomes
much larger and could lead to an interplay between the
dimensionality of niche space and the species richness of the
community.
As a first step to addressing these interesting questions, it is necessary
to develop tractable models of niche space evolution in which
a more realistic, multi-dimensional representation of the niche space
can be incorporated. However, due to their inherent computational
complexity, methods have not been available for solving dynamical
models of the evolution of niche structure under the assumption that
the niche space itself can be multidimensional. An exception is given by a class
of trait-based models, recently proposed to explain plant biodiversity (see~\cite{shipley2006plant, laughlin2012predictive, laughlin2014applying}). Although these models are able to incoporate multiple traits, they do not contain
any dynamical element, so that it is not possible to infere which ecological
driver is responsible for niche diversification.

The purpose of this paper is develop and solve a framework for
analyzing ecological models in which the niche space is
multi-dimensional. We represent ecological niches by sequences, and
model competition using the Hamming distance between two sequences.
We report analytical progress by introducing a novel transform that diagonalizes the interaction community matrix (i.e. the linear stability operator), and allows us to compute the conditions for niche differentiation and the final individual distributions. Our approach generalizes a previous
study~\cite{rogers2012spontaneous} where competing binary genomes have
been analyzed in a similar fashion.
Unlike~\cite{rogers2012spontaneous}, our analysis is not restricted to
binary sequences, as we consider alphabets of arbitrary size, the size
being different for different symbols in the sequence. Our calculations
are presented in the framework of the competitive Lotka-Volterra model,
following~\cite{pigolotti2007species}, but the results are applicable
to any pattern forming system in sequence space.

This paper is organized as follows. In Sec.~\ref{moddef}, we define the multi-dimensional niche space using sequences, and introduce the Lotka-Volterra dynamics on this space. The mathematical tools for analyzing the model are introduced in  Sec.~\ref{analysis}, with the technical proofs given in Appendix~\ref{A} and Appendix~\ref{B}.
In Sec.~\ref{results_of_analysis} we present specific results for the model under study.
Two cases are discussed: (A) when there is only one unstable pattern
forming mode (non-degenerate case), and (B) when there are many competing, equally unstable, modes (degenerate case).
The non-degenerate case has been reported previously in simpler niche space models~\cite{fort2009paradox,fort2010clumping, maruvka2006nonlocal}. In Sec.~\ref{exp_comp}, we explain how our theory can be used to test ecological hypothesis against experimental data.

The $C$ code we have used is available as online Supplementary Material, where we also provide a Mathematica notebook file with the details of the calculations.

\section{Formulation of trait-based ecological models}\label{moddef}

\subsection{The definition of the niche space}
In our framework, ecological niches are represented by sequences of $L$ symbols, each symbol corresponding to a phenotypical trait that can either denote an aspect of the morphology, the behavior, or resource consumption, of a species. The niche space is static, that is, it is not affected by the dynamics of the populations. Moreover, each individual can occupy only a single niche.

To be concrete, we assume that in a hypothetical ecosystem,
individuals are characterized by $L=3$ traits, for example: (i) the
source from which water is collected, (ii) the preferred nesting
place, and (iii) the preferred prey. Each trait $i$ admits a
certain number of options $\Delta_i$. In our example, let us assume
that there are $\Delta_1=3$ water sources (labeled by $W_1$, $W_2$
or $W_3$), $\Delta_2=9$ types of preferred prey ($P_1, \ldots,
P_{9}$), and $\Delta_3=2$ nesting places ($N_1$ or $N_2$). More
generally, the niche space consists of $L$ phenotypic traits with
$\Delta=(\Delta_1, \ldots, \Delta_L)$ possibilities. Each
individual lives in an ecological niche which is denoted by the
letters $I$ or $J$. A niche is obtained by making a choice for each
trait, so that niches are represented by sequences: for instance, a
niche can be $I=W_1N_5P_1$. In niche $I$, there live $n_{I}$
individuals, although we shall use more often the concentrations
$X_{I}$, related to the number of individuals by $n_I= V X_I$.  The
system size is identified in this well-mixed system as the patch
size $V$, and is best thought of as a non-dimensional parameter
controlling the amount of intrinsic noise in the system. Note that
in principle $X_{I}$ can be greater than one. The set of all $X_{I}$
(or $n_{I}$) gives the state of the system.

This way of modeling niches requires a discretization of traits (and their corresponding options) so that it is natural to question whether this is ecologically sound. In fact, some traits, like body size, humidity, altitude or temperature, are better described by continuous variables. However, since the traits are used to distinguish between ecological niches, even the continuous traits need to be binned in order to avoid placing two individuals in different niches due to a negligible difference. For instance, if we introduce the body size in our model, then we classify individuals into various categories such as small, medium, or large body size. If instead we considered the variable continuous, we would treat two individuals of comparable sizes as living in different niches, which is undesirable.

Competition between two individuals depends on the Hamming distance
between the niches in which they live, $d(I,J)$, that is, the number
of positions at which the corresponding symbols are different in the
corresponding sequences~\cite{rogers2012spontaneous}. For example,
the niches $I=W_1N_5P_1$ and $J=W_2N_5P_0$ have Hamming distance
$d(I,J)=2$. Thus, the smaller the Hamming distance, the more two
individuals compete. This means, for example, that if several water
sources are present, we expect the individuals to spread among all
sources but compete only if they collect water from the same source.
Also competition may occur at multiple positions in the sequence, so
that individuals may compete both for a shared water source and a
shared prey. This way of measuring competition depends on
how different two sequences are but it does not matter which trait
is different: two identical niches but with different water sources
have the same Hamming distance of two identical niches but with a
different preferred prey. This issue can be solved by adopting
a more general distance, in which traits are weighted according to a weight
vector, $W = (w_{1},\ldots,w_{L})$. This distance reads:
\be\label{gendist}
d(I,J) = \sum_{l=1}^{L} w_{l} \, \delta_{I_{l},J_{l}},
\ee{}
where $I_{l}$ denotes the $l$-th symbol of niche $I$ ($J_{l}$ is analogous). If $w_{l}=1$ for every $l$, this distance reduces to the Hamming distance. The method presented in Sec~\ref{analysis} can be used with both the Hamming distance and distance~\eqref{gendist}, but the former has been chosen for simplicity.

\subsection{Introducing the dynamics: the Lotka-Volterra equations}
Having defined a niche space, we now need to specify how the number of individuals per niche evolve in time. For simplicity, we follow previous studies~\cite{pigolotti2007species, hernandez2009species, pigolotti2010gaussian} and adopt the Lotka-Volterra equations with an exponential competition kernel, as defined below.

We consider the following equations: \be \label{LV}
    \dot X_I = X_I \left(1 - \frac{1}{C} \sum_J \mathcal G_{IJ} X_J\right),
\ee which model birth and death of organisms via competition. The sum
$\sum_{J}$ is over all possible niches. In this way, we
account for competition between different niches (when $I\neq J$) and
competition within the same niche (for $I=J$). We consider the family
of competition kernels~\cite{pigolotti2007species}
\be
    \mathcal G_{IJ} = \exp\left[-\left(\frac{d(I,J)}{R} \right)^\sigma \right].
\ee The competition length, $R$, and the exponent, $\sigma$, are
positive integers and allow us to consider different choices of the
competition kernel. However, in all of them competition is more fierce
as the Hamming distance decreases since $\mathcal G_{IJ}$ is always decreasing in $d$.
Increasing $\sigma$ stretches the
shape of the competition kernel and as $\sigma\to\infty$, the kernel
$\mathcal G$ tends to a stepwise function. We define the carrying
capacity \be{}\label{eq:defofC}
    C=\sum_J \mathcal G_{IJ},
\ee so that the system admits the fixed point $X^*=1$ in addition to
the fixed point, $X_0=0$, corresponding to mass extinction. The choice
of a constant carrying capacity is consistent, since $\sum_J \mathcal
G_{IJ}$ is independent of $I$; to see this, note that every row in
$\mathcal G$ must be a permutation of another row and thus has the same
sum. The reader may note that choosing the carrying capacity
of the system is unphysical, however, this has the only effect of
renormalizing the fixed point value. Indeed, the same effect is
obtained by rescaling the concentration vector by $X \mapsto A X$ where
$A=C/\sum_J\mathcal G_{IJ}$.

In order to take into account  the effects of  intrinsic
noise~\cite{van1992stochastic}, we define a stochastic model
corresponding to Eq.~\eqref{LV}, using the following transition rates,
$T$, which define the probability per unit of time that birth and
death occur for an individual living in niche $I$: \be\label{stoch}
    \begin{split}
    T(n_I + 1 | n_I) &= X_I, \quad \text{(birth)}\\
    T(n_I - 1 | n_I) &= C^{-1}\sum_J \mathcal G_{IJ} X_I X_J, \quad\text{(death)}.
    \end{split}
\ee The first equation indicates that the number of individuals can
increase by one unit with a probability per unit of time $X_{I}$. The
second equation has an analogous meaning. Note that in the stochastic
model, the quantities $n_{I}$ (and thus $X_{I}$) are subject to
discrete increments, whereas in Eqs.~\eqref{LV} the concentrations are
continuous variables. The difference between the two models is
controlled by $V$ and, as $V\to\infty$, the stochastic
system~\eqref{stoch} recovers the deterministic description in
Eqs.~\eqref{LV}.  All numerical simulations in the paper are performed
using the Gillespie algorithm~\cite{Gillespie1977}, which simulates the
stochastic model~\eqref{stoch}.

\section{Analysis} \label{analysis}
In this Section, we show that the fixed point $X^{*}$ undergoes a
pattern instability in niche space which drives the system to
diversification. To analyze the instability, we define a suitable
transform (Eq.~\eqref{F} below) that diagonalizes matrices whose
element depends on the sequences only via their Hamming distance (i.e.
Hamming matrices). By doing so, we are able to diagonalize the
linear stability operator (i.e. the Jacobian matrix) of the
fixed point $X^{*}$.

Transform \eqref{F} is at the core of our analytical treatment. We have
arrived at this formula by generalizing the Hadamard transform,
$(-1)^{|I \cdot J|}$, previously used in the study of competing binary
genomes~\cite{rogers2012spontaneous}. Another way to understand
Transform~\eqref{F}, is by noting that Hamming matrices are special
cases of a general class of  matrices called {\it block circulant with
circulant blocks} (BCCB), whose diagonalizer is
known~\cite{davis1979circulant}. Using this latter fact, we show that
the spectrum of Hamming matrices can be obtained explicitly
(Eq.~\eqref{spJ}), which allows a straightforward investigation of the
properties of the pattern instability.

\subsection{Pattern instability in niche space}
The Lotka-Volterra equations~\eqref{LV} admit the fixed point,
$X^{*}=1$, which corresponds to a homogeneous distribution of
individuals in niche space. If the fixed point is unstable, small
perturbations grow exponentially fast and the system relaxes to a
non-homogeneous profile, as described by pattern formation
theory~\cite{cross2009pattern}. To inspect for instabilities, we
linearize Eq.~\eqref{LV} around the fixed point $X^*$. Denoting the
small deviations by $\delta X_I = X_I - X^*$, we arrive at
(in vectorial notation): \be \label{xlin}
    \frac{d}{dt}{\delta X} = \mathcal J \delta X = - \frac{1}{C}\mathcal G\, \delta X.
\ee To check for the stability of this system, we diagonalize the
linear stability operator $\mathcal J$.  Again recall that
the elements of $\mathcal J$ are defined by \be\label{f}
    \mathcal J_{IJ} = f(d(I,J)),
\ee and retain a dependence in the sequences $I$ and $J$ only via their
Hamming distance. We call matrices with this property \textit{Hamming
matrices}. The overall dimension of matrix $\mathcal J$ is $D \times
D$, where $D$ is the total number of niches, namely, \be\label{D}
    D = \prod_{l=1}^{L}\Delta_l.
\ee{} As shown in the Appendix~\ref{A}, Hamming matrices are special
cases of a general class of circulant matrices and are
diagonalized by \be \label{F}
    \mathcal F = \mathcal F_{\Delta_1} \otimes \ldots \otimes \mathcal F_{\Delta_L},
\ee where the symbol $\otimes$ indicates the Kronecker
product between two matrices $\mathcal A$ and $\mathcal B$:
\be\label{kron_prod}
    \mathcal A \otimes \mathcal B =
    \begin{bmatrix}
        \mathcal A_{11}  \mathcal B & \cdots & \mathcal A_{1d} \mathcal B \\
        \vdots & \ddots & \vdots \\
        \mathcal A_{d1} \mathcal B & \cdots & \mathcal A_{dd} \mathcal B
    \end{bmatrix}.
\ee The matrix $\mathcal F_{\Delta_l}$ ($l=1,\ldots,L$) is the
$\Delta_{l}\times\Delta_{l}$ discrete Fourier matrix defined by:
\be
    (\mathcal F_{\Delta_l})_{jk} = \exp\left(i \frac{2 \pi j k}{\Delta_{l}} \right)
\ee where $i$ denotes the imaginary unit and with normalization
$\mathcal F_{\Delta_l}^\dagger \mathcal F_{\Delta_l} = \mathbb
I_{\Delta_{l}}$ (the symbol $\dagger$ stands for the conjugate
transpose and $\mathbb I_{\Delta_{l}}$ is the $\Delta_{l}$
-dimensional identity matrix). The indexes $j$ and $k$ range from
one to $\Delta_{l}$. Since $F_{\Delta_{l}}$ is unitary, then
$\mathcal F$ is unitary as well: $\mathcal F^\dagger \mathcal F =
\mathbb I_D$. Note that Transform $\mathcal F$ is not, in general, a
Fourier transform, since (for example for $L=2$ and
$\Delta_{1}=\Delta_{2}=2$), $\mathcal F = \mathcal F_2 \otimes
\mathcal F_2 \neq \mathcal F_4$.

Applying $\mathcal F$ to both sides of Eq.~\eqref{xlin} yields the
decoupled equations \be{}
     \frac{d}{dt}{\delta\widetilde{ X_K}} = J_K  \widetilde {\delta X_K},
\ee
where we have defined the transformed vector,
\be{}
    \delta \widetilde {X_K} = \sum_I \mathcal F_{KI}\delta X_I,
\ee
and $J_{K}$ are the eigenvalues of matrix $\mathcal J$:
\be
    J_K = (\mathcal F \mathcal J \mathcal F^\dagger)_{KK}.
\ee Note that the eigenvalues $J_{K}$ are real given that Hamming
matrices are symmetric, i.e. $\mathcal J_{IJ} = \mathcal J_{JI}$.

The variable $K$ is the conjugate variable, in transformed space, to
the sequence variable $I$, and ranges from one to $D$. When
$J_K>0$, for some $K$, the fixed point $X^{*}$ is unstable and the
amplitude of the corresponding eigenmode $v^{(K)}$, with component
$v^{(K)}_M = \mathcal F^\dagger_{MK}$, grows in the system. Note that
since $\mathcal F$ is unitary, its rows form an orthonormal
basis in sequence space, with respect to the canonical scalar product
so that \be \label{scprod}
    v^{(K_{1})} \cdot v^{(K_{2})} = \sum_{M=1}^{D}  v^{(K_{1})}_M v^{*(K_{2})}_M = \delta_{K_{1},K_{2}}.
\ee
The symbol $*$ stands for the complex conjugate and the dimension
$D$ is given by Eq.~\eqref{D}.

\subsection{Spectrum of Hamming matrices}\label{sp_sec}
Transform~\eqref{F} diagonalizes matrix~\eqref{f}, yet, carrying out
the matrix product $\mathcal F \mathcal J \mathcal F^\dagger$ may not
be feasible as the dimensionality $D$ of Hamming matrices can be very
large, even for low dimensional niche spaces. However,
by extending Theorem 5.8.1 of~\cite{davis1979circulant}, we can obtain
a compact expression for the eigenvalues and eigenmodes of any Hamming
matrix $\mathcal J$. The result is stated in the following and proved
in Appendix~\ref{A}.

Let us consider any Hamming matrix, $\mathcal J$, defined by
Eq.~\eqref{f} via a certain function $f$. Then the $K$-th
eigenvalue of $\mathcal J$, denoted by $J_K$, is given by
\be
\label{spJ}
    \begin{split}
    & J_K =\left(\prod_{l=1}^{L} \sum_{k_l=0}^{\Delta_l-1} \left( B(k_1,\ldots,k_L) \left( \Omega_1^{k_1} \otimes \ldots \otimes \Omega_L^{k_L} \right) \right)\right)_{KK},
    \end{split}
\ee
where we have used the following definitions:
\be \label{def}
    \begin{split}
    B(k_1,\ldots,k_L) &= f(L-\sum_{l=1}^{L} \delta_{k_l,0}), \\
    \omega_l = \exp\left(i \frac{2 \pi}{\Delta_l}\right), \quad \Omega_l &= \left(1,\omega_l, \omega_l^2, \ldots, \omega_l^{\Delta_l-1} \right).
    \end{split}
\ee The dependence on matrix $\mathcal J$ is contained in function $B$,
which returns the function $f$, evaluated on the number of non-zeros
which are passed to $B$ as argument. For example, for $L=3$,
$B(1,0,0)=f(2)$. In Eq.~\eqref{spJ}, the notation $\Omega_1^{k_1}$ means that every element of the vector $\Omega_{1}$ is elevated to the power $k_{1}$. Also, note that $J_{K}$ is a scalar quantity, given by the $K$-th entry of the diagonal of the matrix defined between parentheses in the RHS of Eq.~\eqref{spJ}. The supplementary Mathematica file contains an implementation of this formula.

The corresponding eigenmode to the $K$-th eigenvalue reads:
\be\label{eigenmodes}
    v^{(K)}_M = \mathcal F^\dagger_{MK}.
\ee
Thus, the eigenmodes do not depend on the system parameters but only on the dimension of the niche space. This fact is not surprising, as it is analogous to what occurs in other cases, such as in systems diagonalized by a discrete Fourier transform.

Note that if an eigenvalue is degenerate, i.e. there is more than
one corresponding eigenmode, then the eigenvalue appears in
Eq.~\eqref{spJ} once for each eigenmode.

\begin{figure}[htpc!]
\includegraphics[width = 0.95\columnwidth]{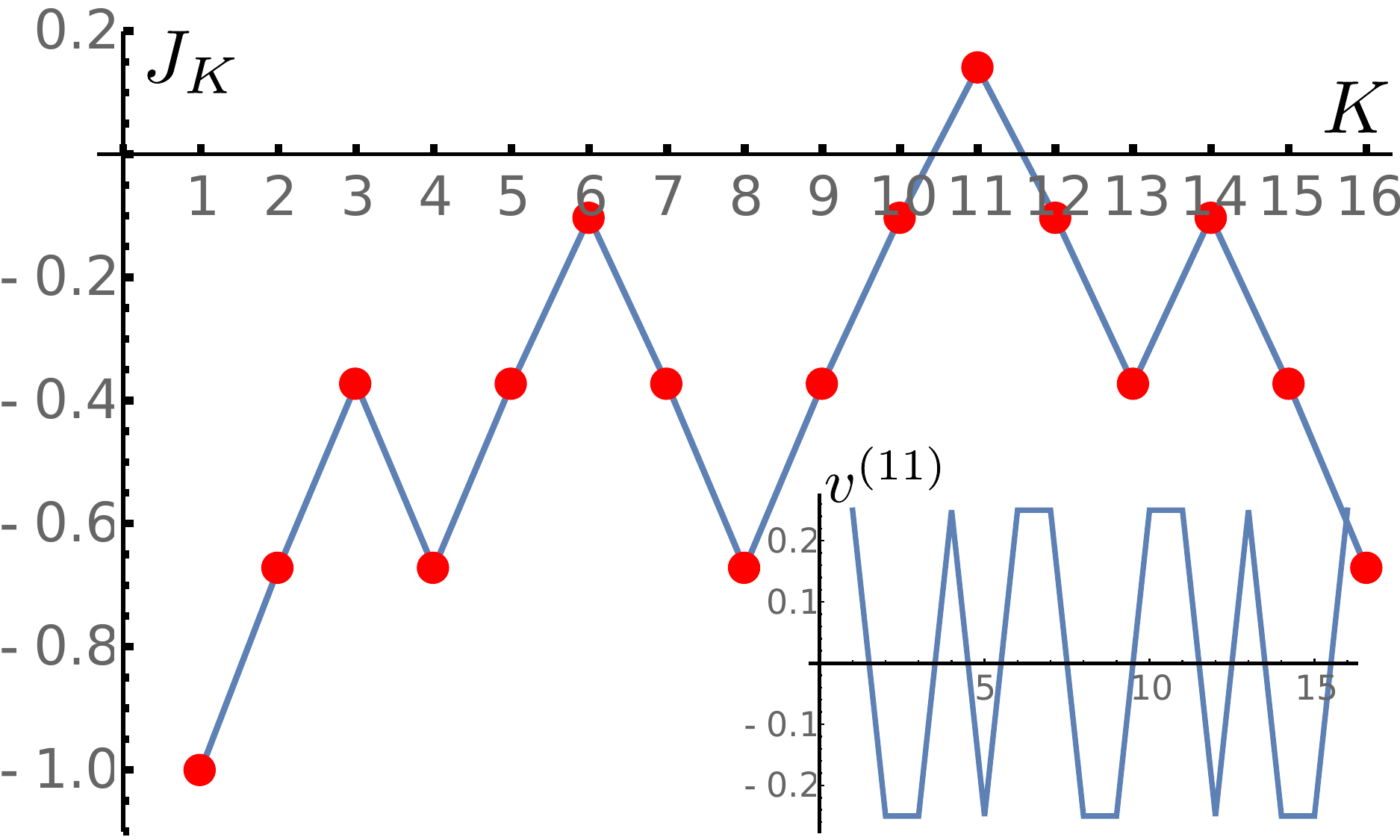} \\
\caption{(Color online) (Main figure) The spectrum of $\mathcal J$, $\{J_K\}$ (red
dots), is calculated with Eq.~\eqref{spJ} and shown as a function of
the increasing wavelength $K$, after a reordering as explained in
Sec.~\ref{nondeg_inst}. The blue line has been added for clarity.
Parameter values: $L=4$, $\Delta_l=2$, $R=1$ and $\sigma=2$. (Inset)
The components of eigenmode $v^{(11)}$ for the same parameter values.
\label{fig1}}
\end{figure}

In the following section, we use formulae~\eqref{spJ} and~\eqref{eigenmodes} for computing eigevalues and eigenmodes of Hamming matrices. There is an alternative way for representing their spectrum, which is given, and proved, in Appendix~\ref{B} (Eq.~\eqref{eq:defoflambda}). In doing that, we have been able to show that the leading eigenvalue in pattern-forming instabilities is non-degenerate if and only if the alphabet is binary (i.e. $\Delta_{l}=2$, for each $l$). Thus, the typical case consists of a degenerate instability, where many equally unstable eigenmodes compete for their emergence. In the next Section, we discuss how the linear theory can be used to predict the asymptotically long time population distributions of niches.

\section{Results}\label{results_of_analysis}
We have established the mathematical tools that we need to inspect
the pattern instability in Eqs.~\eqref{LV}. We now show that the
linearly unstable eigenmodes give a prediction for the final
distribution of individuals in niche space. Two cases are investigated:
a non-degenerate and degenerate instability. In the former case the
analytical prediction matches the result of stochastic simulations. In
the latter, simulations show a different final individual distribution
at every run, due to stochastic effects (or the initial condition) that
randomly privilege some of the equally unstable competing eigenmodes.
However, the dynamics averaged over many run shows consistency with the
prediction of the linear theory, as also reported previously in a one-dimensional niche
model~\cite{fort2009paradox,fort2010clumping}.

In Sec.~\ref{deg_inst}, we also report the observation of
stochastic patterns, or noise-induced patterns, which arise when a
weakly stable eigenmode is subject to noise. Since we investigate cases
where the homogeneous state is linearly unstable, stochastic patterns
are superposed to deterministic patterns, and the difference between
the two is that the amplitude of stochastic patterns decrease as the patch size
$V$ increases.

\subsection{Case: Non-degenerate instability}\label{nondeg_inst}
We first study a simple case that displays a single mode instability.
We begin by considering binary sequences of four bits, $L=4$,
$\Delta_l=2$ (for every $l=1,\ldots,4$), and parameter values $R=1$ and
$\sigma=2$. We find it useful to reorder the eigenmodes so that we can
interpret $K$ as a wavelength. For the case of binary sequences,
the eigenmodes~\eqref{eigenmodes} are manifestly real and we
define the wavelength as the number of times that the eigenmode
$v^{(K)}$ crosses the $K$ axis. We then reorder the eigenmodes by
increasing wavelength. Note that is always possible to choose
a real basis of eigenmodes given that Hamming matrices are symmetric,
however, since we use Eq.~\eqref{eigenmodes} for their expressions, the
chosen eigenmodes are real only for certain cases.

The spectrum of matrix $\mathcal J$, calculated using Eq.~\eqref{spJ}
and then reordered, is shown in Fig~\ref{fig1}. Each eigenvalue (red
dots) is stable except the one with wavelength $K=11$
($K=16$, without reordering) and the profile of the
corresponding eigenmode is shown in the inset. Thus, starting close to
the homogeneous state causes the growth of the eigenmode $v^{(11)}$
whilst the other eigenmodes decay away. The growth is eventually damped
by the effect of the non-linearities, which become relevant as the
system moves away from the homogeneous state.

\begin{figure}[htpc!]
\includegraphics[width = 0.95\columnwidth]{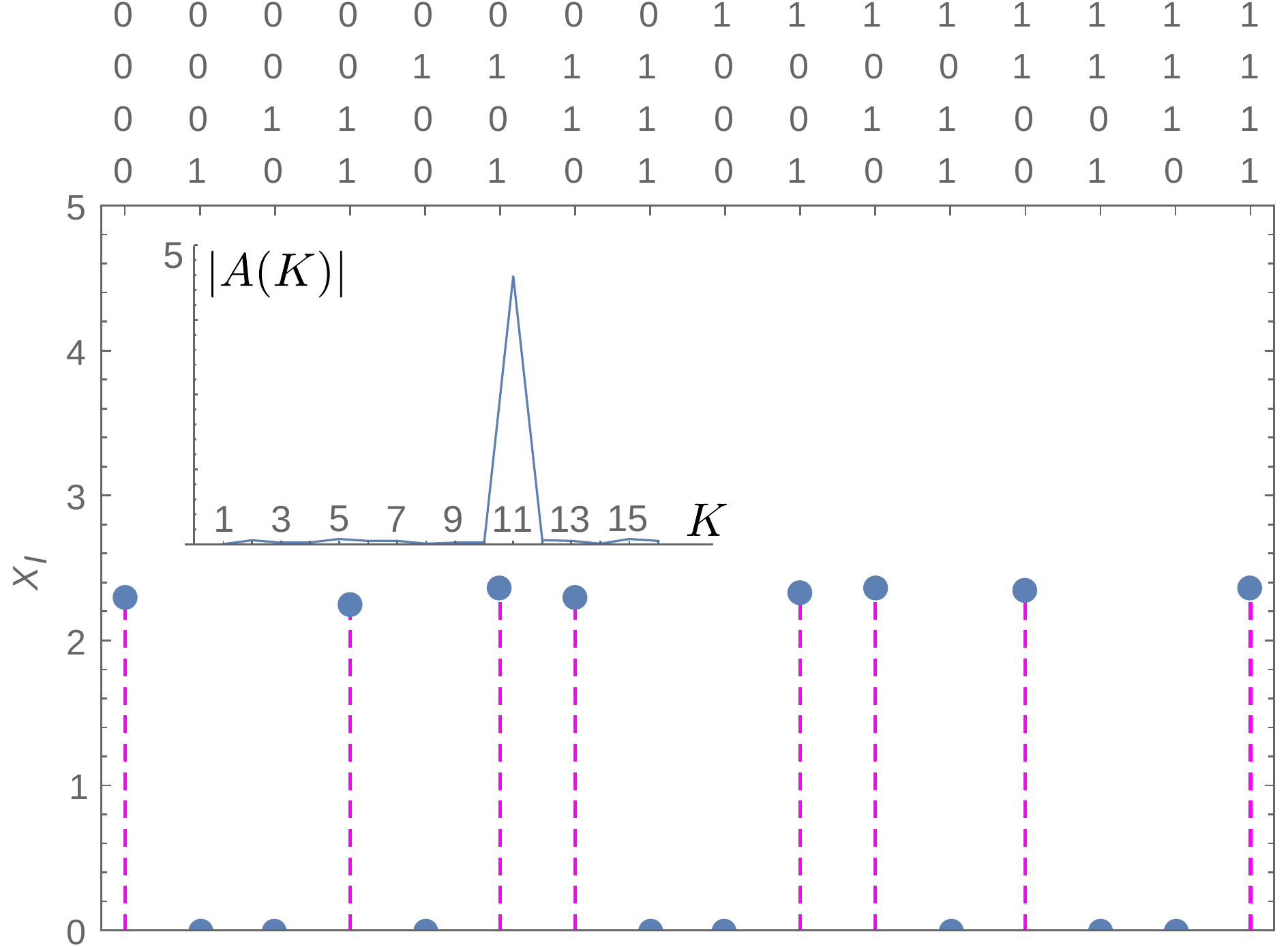} \\
\caption{(Color online) (Main figure) The concentrations $X_I$ as a function of the
individuals $I$ (represented as binary sequences at the top of the
figure), obtained by running stochastic simulations for time
$t=V*10^{2}$ with $V=10^{3}$ and other parameter values as in
Fig.~\ref{fig1}. (Inset) The absolute value of $A(K)$, Eq.~\eqref{AK},
obtained from the final individual distribution shown in the main
figure.  \label{fig2}}
\end{figure}

The final individual distribution is given by a superposition of those
eigenmodes predicted to be unstable in the linear analysis. In this case,
there is a single unstable mode so that we expect the final individual
distribution to exhibit a shape analogous to $v^{(11)}$.  The results
of stochastic simulations, displayed in Fig.~\ref{fig2}, show agreement
between the final state and that predicted in the inset of
Fig.~\ref{fig1}. Note however, that on some simulation run, the pattern
may sometimes appear reversed, as either of the eigenmodes $v^{(11)}$
or $-v^{(11)}$ may grow.

Another way to check the agreement between theory and simulations
consists of expressing the final state $X$, measured from the
simulations, as a superposition of the eigenmodes, i.e. \be{}
    X_{I} - \bar X =\sum_{K}A(K)v_{I}^{(K)}
\ee The profile $X$ is renormalized to zero-average, by subtracting
$\bar X = d^{-1} \sum_{I} X_{I}$, in order to avoid a large component
in $A(K=0)$. The quantities $A(K)$, obtained by taking the scalar
product between $X$ and $v^{(K)}$, give the extent to which that
eigenmode emerges in the final pattern. Specifically:
\be \label{AK}
A(K) =  \left( X-\bar X\right) \cdot v^{(K)},
\ee{}
where the dot is the scalar product~\eqref{scprod}. For our previous
case, we expect that $A(K)$ is approximately zero for every
wavelength except $K=11$, on which it takes a positive (resp.
negative) value, given that $v^{(11)}$ (resp. $-v^{(11)}$) has
grown. This is confirmed by the inset of Fig.~\ref{fig2}, where the
absolute value of $A(K)$ is shown. Note that both
solutions, $v^{(11)}$ and $-v^{(11)}$, yield the same contribution
to $|A(K)|$, since we have taken the absolute value.

The solution displayed in Fig.~\ref{fig2} is metastable, in that, sooner or later a large rare fluctuation will lead the system to extinction, since $n_{I} = 0$, for all $I$, is an absorbing state. Such a fluctuation is very rare and not observable in simulations for timescales $\sim V*\dot 10^{6}$. Therefore, the solution can be considered evolutionarily stable for practical purposes.

As a final remark, let us note that we have verified the
agreement between linear theory and simulations for various parameter
instances and noise realizations. Results are not shown for the sake of
compactness, but the case discussed above is the prototypical example
when a single-mode instability is in play.

\subsection{Case: Degenerate instability} \label{deg_inst}
\begin{figure}[htpc!]
\includegraphics[width = 0.95\columnwidth]{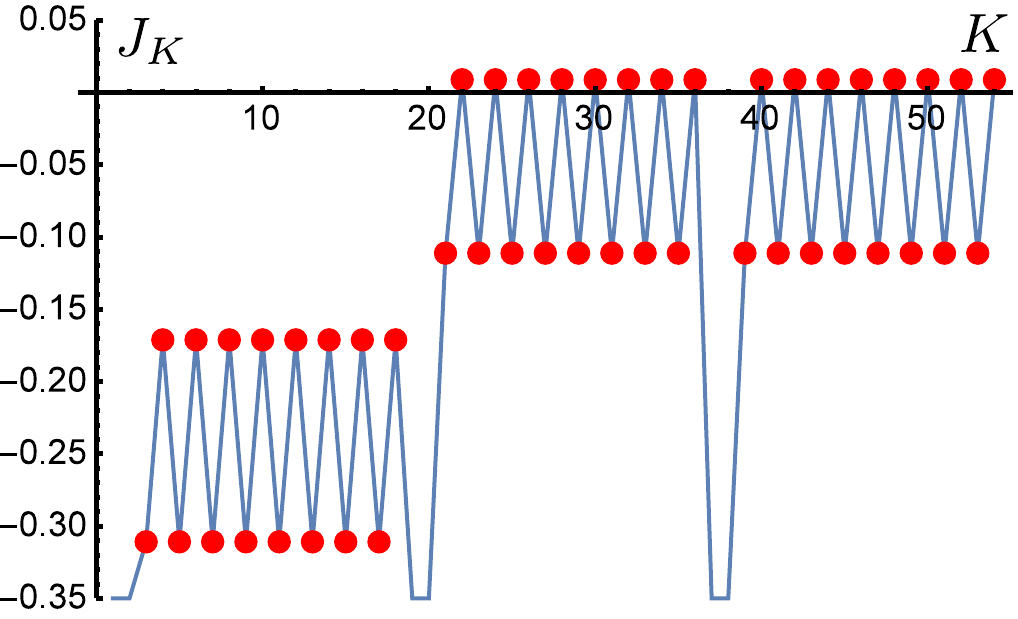} \\
\caption{(Color online) The spectrum of $\mathcal J$, $\{J_K\}$ (red dots), is
calculated with Eq.~\eqref{spJ} and shown as a function of $K$ for
parameter values: $L=3$, $\Delta_{1}=3$, $\Delta_{2}=9$ and
$\Delta_{3}=2$ with $\sigma=2$ and $R=1$. The blue line has been added
for clarity. \label{fig3}}
\end{figure}

Does the linear stability analysis provide a reliable prediction for a
general case? Typically the instabilities in these kinds of models are
highly degenerate as they possess many equally unstable eigenmodes. For
example, let us consider the case discussed in the Introduction where
$L=3$, $\Delta_{1}=3$, $\Delta_{2}=9$ and $\Delta_{3}=2$ with
$\sigma=2$ and $R=1$. For this case we do not reorder the eigenmodes.
The spectrum (dots of Fig.~\ref{fig3}) indicates that there are sixteen
unstable eigenmodes, each corresponding to the same eigenvalue. When a
degenerate instability is in play, the fate between the competing
eigenmodes is determined by the non-linearity, the intrinsic noise and
the initial condition so that deviations from the linear prediction are
expected.

\begin{figure}[htpc!]
\includegraphics[width = 0.95\columnwidth]{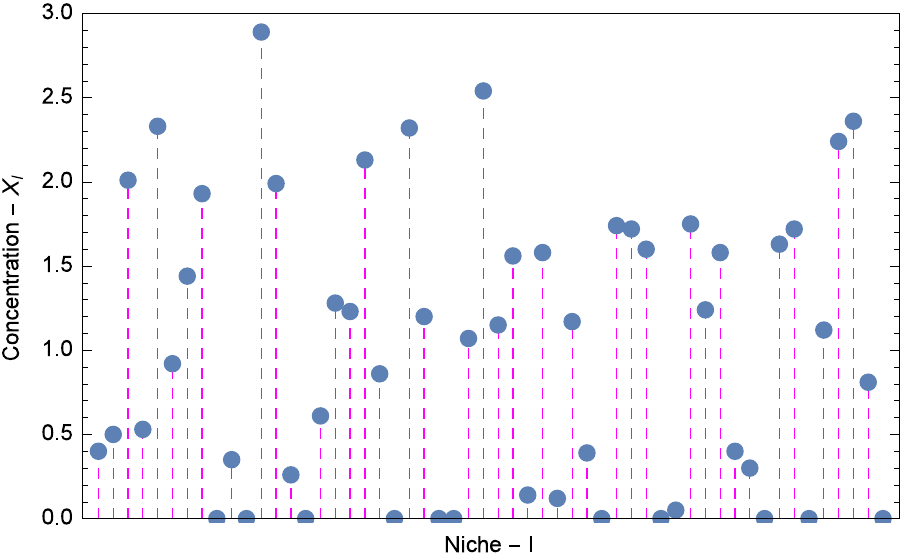} \\
\caption{(Color online) Final niche distribution of a single run. Parameter
values as in Fig.~\ref{fig3} but with $V=10^{2}$. The simulation has
run for time $T=10^{3}\cdot V$. \label{fig4}}
\end{figure}

For example, let us examine the result of a single run, shown
in Fig.~\ref{fig4}. Unlike the binary alphabet case, the
behavior displayed is now quite rich: in some niches the population goes
extinct, while other niches are scarcely populated, and few of them contain a
large number of individuals. Computing the profile  $A(K)$ for this
niche distribution (not shown), yields a significantly different result
to what predicted in Fig.~\ref{fig3}. Since there are many equally
unstable eigenmodes, stochasticity gives a random advantage to some of them,
which then grow faster and overwhelm the growth of the other unstable
modes. Only the eigenmodes which have been privileged in this way
appear in the final profile, which is thus not predictable.

\begin{figure}[htpc!]
\includegraphics[width = 0.95\columnwidth]{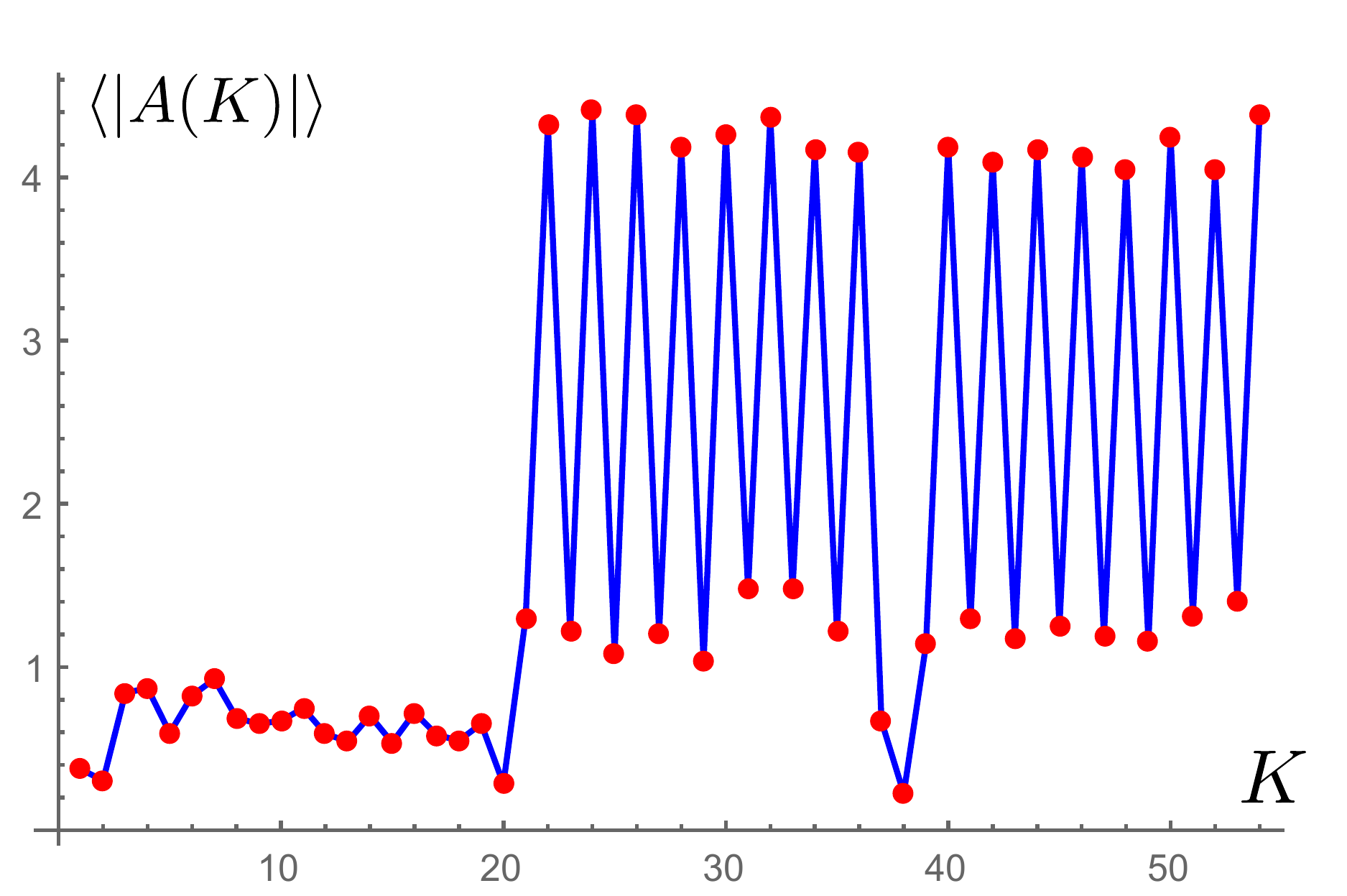} \\
\caption{(Color online) The average absolute value of $A(K)$ (red dots),
obtained by averaging formula~\eqref{AK} over $10^{4}$ runs. Parameter
values as in Fig.~\ref{fig3} but with $V=10^{2}$. Each simulation has
run for time $T=10^{3}\cdot V$. The blue line has been added for
clarity. \label{fig5}}
\end{figure}

Although the fate of a single run is not captured by our
analysis, we may ask whether the average behavior resembles the
prediction of the theory. We therefore compute the profile $|A(K)|$,
averaged over several runs. The result, shown in Fig.~\ref{fig5},
indicates that the highest values of $\langle|A(K)|\rangle$ correspond
indeed to the eigenmodes predicted to be unstable by the spectrum in
Fig.~\ref{fig3}. Interestingly, the profile $|A(K)|$ assumes small, but
non-zero values, for the $K$s corresponding to stable eigenmodes. This
is an example of stochastic patterning --- pattern formation caused by
a slowly relaxing eigenmode subject to intrinsic noise --- which have
already been observed in predator-prey~\cite{butler2009robust} and
reaction-diffusion systems~\cite{biancalani2010stochastic}.

Stochastic patterns could have also been visible in the case
studied in the previous section since, as shown in Fig.~\ref{fig1}, the
wavelengths $K=6$ and $K=14$ are close to the onset of instability.
However, unlike deterministic pattern formation, the amplitude of
stochastic patterns depends on the magnitude of the perturbation which
cause them, the intrinsic noise, and therefore scales as
$V^{-1/2}$~\cite{van1992stochastic}. Having chosen $V=10^{3}$ for
generating Figs.~\ref{fig1} and~\ref{fig2}, and $V=10^{2}$ for
Figs.~\ref{fig3} and~\ref{fig5}, has rendered the stochastic patterns
visible only in the latter two figures. Indeed, redoing the simulations
for Fig.~\ref{fig5} but with a larger value for $V$, yields a profile
$\langle|A(K)|\rangle$ which is zero everywhere except the unstable
eigenmodes.

\section{Comparison against experimental data} \label{exp_comp}
The theory presented so far provides a tool for linking niche distributions, based on multiple traits, to the individual-based interactions that govern ecosystems. Our approach can also be used to test an ecological idea against data from a real ecosystem. As a prototypical example, let us consider a certain dataset representing species abundances and ask the following question: ``Is competitive exclusion the primary driver of ecological diversification?''. The series of steps below instructs how to use our theory to attack this problem.


\subsubsection{Representing ecological and/or morphological niches}
To begin with, we extrapolate the niche distribution from the species abundances data. In order to identify the possible niches, we need to select those traits within which we want to divide the species. These traits can refer both to the morphology of the species  (e.g. the body size) or describe an aspect of their behavior (e.g. the nesting place). We allow a number of possibilities to each trait (e.g. individuals of small, medium or large body size), so that the niche space can be mapped into the abstract sequence space, in the same way as described at the beginning of Sec.~\ref{moddef}. We can then visualize the number of individuals per niche, in a similar fashion to Fig.~\ref{fig4}.

\subsubsection{Transforming the niche distribution}
Once the sequence space has been established, we can define Transform~\eqref{F}, that relies solely on the geometry of the niche space (i.e. how many traits / how many possibilities per trait). We then transform the niche distribution, using the eigenmodes~\eqref{eigenmodes} and Eq.~\eqref{AK}, to yield a figure analogous to Fig.~\ref{fig5}. The peaks of the figure will highlight the emerging eigenmodes in the system. The aim is to predict the emergence of those peaks starting from the individual-based interactions.

\subsubsection{Agreement with an individual-based model}
The individual-based model are represented by set of differential equations such as System~\eqref{LV}, and should include the effects which are supposedly the main drivers for niche diversification. For example, if we assume that competitive exclusion is the sole, or principal, driver, then we can use directly the Lotka-Volterra equations~\eqref{LV}. The prediction of the model can be read off by looking at its spectrum, such as in Fig.\ref{fig3}: The eigenmodes predicted unstable are those which correspond to a positive eigenvalues, that is, appear above the $K$ axis. These eigenmodes should be compared against those observed in the data. If an agreement between the two sets of eigenmodes indicates that the interactions in the individual-based model are effectively responsible for niche diversification.

\section{Conclusion}
In this paper, we have proposed a class of ecological models
which displays niche diversification due to competitive interactions.
The novelty of the work lies in the fact that the niche space is
high-dimensional: An arbitrary number of phenotypic traits can be
included in the model, each admitting a certain number of
possibilities. In this way, the geometry of the niche space is
intrinsically different from the Huchinsonian picture where ecological
niches are represented by hypercubes in $\mathbb R^n$. Also, niche
overlapping is given by the Hamming distance, rather than the Euclidean
distance, which is a simpler and a more realistic way to quantify how
many traits two individuals have in common.

Most of the paper is centered on the mathematical aspects for
analyzing these models. We have shown that the underlying pattern
instability can be predicted using the mathematical tools presented in
Sec.~\ref{sp_sec}. The linear stability operators (i.e. Jacobian
matrices) depends on the niches only via their Hamming distance, and we
have called such matrices \textit{Hamming matrices}. We have shown that
Hamming matrices are special cases of class of circulant matrices (i.e. BCCB), so
that we can use the diagonalizer of the latter for obtaining the
spectrum of Hamming matrices. In this way, we have arrived at
formulas~\eqref{spJ} and~\eqref{eigenmodes}. By using these
expressions, we have shown that the linear stability analysis predicts
the final individual distribution in niche space, although the
prediction works only on average if a degenerate instability is
present.

An alternative interpretation for our class of models is that of
interacting genomes, where $L$ represents the genome length and
$\Delta_l=4$ for all $l$ (the symbols are now nucleotides: $A$, $G$,
$T$, $D$). Indeed, the advent of population genetics has stimulated
a similar kind of modeling, which has attracted interest in the
physics community thanks to quasispecies theories~\cite
{nilsson2000error, altmeyer2001error, cohen2005recombination,
park2007phase, jain2007adaptation} and paramuse models of evolution~
\cite{crow1970introduction, jain2007adaptation}. The connection
between these studies and our work, is that in both cases the models
are sequence-based~\cite{jain2007adaptation}, in that they describe
the dynamics of an interacting population in which each individual
is represented by a sequence. Binary sequences are often analyzed as
they allow analytical approaches, such as mapping the model into the
Ising model~\cite{baake1997ising, baake2001mutation,
saakian2006exact}, considering various limits~\cite
{peliti2002quasispecies} or by using an Hadamard transform~\cite
{rogers2012spontaneous}. We expect our theory to be applicable to
these models as well, allowing for generalizations in which larger
alphabets are considered. This represents a possible direction for
future works.

Finally, let us notice that throughout the paper, our
analytical treatment is limited to the deterministic level for
simplicity, even though we have shown that  the stochastic model
exhibits stochastic patterning, caused by weakly stable eigenmodes
subject to intrinsic noise, as shown in~ \cite
{rogers2012spontaneous} for the case of binary sequences. The same
authors have also reported a different type of stochastic patterning
based on the multiplicative nature of noise~ \cite
{rogers2012demographic}, which occurs where the noise is strong in
the system. This effect can lead to stable, noise-induced patterns.
We expect that our model can exhibit this type of order as well.
Further investigations will be devoted to extend our method for
analyzing the stochastic counterparts of these models.

\vspace{10pt}
\begin{acknowledgments}
T.B. owes much to Elbert Branscomb and Louise Dyson for inspiring
discussions and support received during the time the research has
been carried out. This work has highly benefited from the critical
help of Axel G. Rossberg and Luis Fernandez Lafuerza. This work was
partially supported by the National Aeronautics and Space
Administration through the NASA Astrobiology Institute under
Cooperative Agreement No. NNA13AA91A issued through the Science
Mission Directorate.
\end{acknowledgments}
\bibliographystyle{apsrev4-1}
\bibliography{literature}
\begin{appendix}
\section{Block circulant matrices with circulant blocks}\label{A}

The aim of this Appendix is to prove formulae~\eqref{spJ}
and~\eqref{def} of the main text. These equations provides an
expression for the spectrum of a matrix $\mathcal J$ whose element,
$\mathcal J_{IJ}$, retains a dependence in $I$ and $J$ solely via their
Hamming distance, i.e., $\mathcal J_{IJ} = f(d(I,J))$. We have called
these matrices, \textit{Hamming matrices}. We shall show that these
matrices possess a  block-circulant structure (defined in the
following), which allows us to compute their eigenvalues and
eigenmodes. The reference for this Section is the book of
Davis~\cite{davis1979circulant}.

A matrix is \textit{circulant} if each row vector is rotated one
element to the right relative to the preceding row vector. Clearly, a
circulant matrix is fully specified by one row as the others are simply
given by cyclic permutations. For example, a $3 \times 3$ circulant
matrix has the form: \be \label{J} \mathcal J =    \begin{bmatrix}
                c_{0} & c_{2} & c_{1} \\
                c_{1} & c_{0} & c_{2} \\
                c_{2} & c_{1} & c_{0}
                \end{bmatrix}.
\ee{} In this definition, the symbols $c_{0}, c_{1}$ and $c_{2}$
represent numbers. On the other hand, if $c_{0}, c_{1}$ and $c_{2}$ are
circulant matrices themselves, then $\mathcal J$ is called a
\textit{circulant matrix of level two}. More generally, a
\textit{circulant matrix of level $L$} can be decomposed in blocks
which are circulant matrices of level $L-1$. A circulant matrix of
level one is tantamount to say that the matrix is circulant.

The size of the blocks can be different at each step and specifies the
\textit{type} of the matrix. For example, we say that a circulant
matrix of level $3$ is of type $(\Delta_{1},\Delta_{2},\Delta_{3})$, if
it can be divided in $\Delta_{1}\times\Delta_{1}$ blocks, each of which
can be divided in $\Delta_{2}\times\Delta_{2}$ blocks, each of which is
a circulant matrix with dimension $\Delta_{3}\times\Delta_{3}$. In
general, a circulant matrix of level $L$ is of type $\Delta=
\left(\Delta_{1},\ldots,\Delta_{L}\right)$. Thus, the level is
specified automatically by the length of the type. The following matrix
is an example of type $(2,3,2)$: \be \label{P} \mathcal P_{(2,3,2)} =
\left[
\begin{array}{cccccccccccc}
 0 & 1 & 1 & 2 & 1 & 2 & 1 & 2 & 2 & 3 & 2 & 3 \\
 1 & 0 & 2 & 1 & 2 & 1 & 2 & 1 & 3 & 2 & 3 & 2 \\
 1 & 2 & 0 & 1 & 1 & 2 & 2 & 3 & 1 & 2 & 2 & 3 \\
 2 & 1 & 1 & 0 & 2 & 1 & 3 & 2 & 2 & 1 & 3 & 2 \\
 1 & 2 & 1 & 2 & 0 & 1 & 2 & 3 & 2 & 3 & 1 & 2 \\
 2 & 1 & 2 & 1 & 1 & 0 & 3 & 2 & 3 & 2 & 2 & 1 \\
 1 & 2 & 2 & 3 & 2 & 3 & 0 & 1 & 1 & 2 & 1 & 2 \\
 2 & 1 & 3 & 2 & 3 & 2 & 1 & 0 & 2 & 1 & 2 & 1 \\
 2 & 3 & 1 & 2 & 2 & 3 & 1 & 2 & 0 & 1 & 1 & 2 \\
 3 & 2 & 2 & 1 & 3 & 2 & 2 & 1 & 1 & 0 & 2 & 1 \\
 2 & 3 & 2 & 3 & 1 & 2 & 1 & 2 & 1 & 2 & 0 & 1 \\
 3 & 2 & 3 & 2 & 2 & 1 & 2 & 1 & 2 & 1 & 1 & 0 \\
\end{array}
\right].
\ee

Let us now consider a niche space defined by sequences, as
explained in Sec.~\ref{moddef}. We assume that sequences are long $L$
characters, the character at position $l$ chosen from an alphabet of
size $\Delta_{l}$. If $\Delta_{l}<10$, for every $l$, we can represent
the sequences using the digits $0\, \textendash\, 9$; e.g., $\Delta_{1}=3$
indicates that at position one of the sequence there is one of the
three symbols: $0$, $1$ or $2$. The size of each alphabet is summarized
by the vector $\Delta= (\Delta_{1},\ldots,\Delta_{L})$ and the
sequences are as ordered as the corresponding numbers. For example, the
sequence space defined by $\Delta = (4,2,3)$ starts from $000$ and ends
in $312$. The number of possible sequences is
$D=\prod_{l=1}^{L}\Delta_l = \Delta_{1}\Delta_{2}\Delta_{3}=24$.

The Hamming distance between two sequences $I$ and $J$ is denoted by
$d(I,J)$ and corresponds to the number of positions at which the
corresponding symbols are different. For instance, $I=102$ and $J=100$
have Hamming distance one. Two sequences are identical if and only if
their Hamming distance is zero.

Let us consider the simplest Hamming matrix of type $\Delta$,
$\mathcal P_{\Delta}$, in which the function $f$ is the identity. The
corresponding matrix element is $\mathcal P_{\Delta, IJ} = d(I,J)$.
Considering, for example, $\Delta=(2,2)$, matrix $\mathcal P_{\Delta}$
looks as follows: \be \label{P22} \mathcal P_{(2,2)} =    \left[
    \begin{array}{cccc}
    0 & 1 & 1 & 2 \\
    1 & 0 & 2 & 1 \\
    1 & 2 & 0 & 1 \\
    2 & 1 & 1 & 0 \\
    \end{array}
    \right].
\ee{}

It is clear that $\mathcal P_{(2,2)}$ is also a circulant matrix of
type $(2,2)$. In general, a Hamming matrix acting on a sequence space
defined by $\Delta$ is a circulant matrix of type $\Delta$.

Any circulant matrix $\mathcal J$ of type $\Delta$ admits a
decomposition (see Theorem 5.8.1 in~\cite{davis1979circulant}) which, for simplicity, is given
in the following for the case $L=2$ (generalization to arbitrary $L$
are straightforward): \be\label{Jdec}
    \mathcal J = \mathcal F^{\dagger} \left( \sum_{k_{1}=0}^{\Delta_{1}-1} \sum_{k_{2}=0}^{\Delta_{2}-1} B(k_{1},k_{2}) \left(\Omega_1^{k_1} \otimes \Omega_2^{k_2}\right) \right) \mathcal F.
\ee{} The decomposition uses the definitions of $\mathcal F$ and
$\Omega_l$ in Eqs.~\eqref{F} and~\eqref{def}. The function
$B(k_{1},k_{2})$ returns the element of $\mathcal J$ corresponding to
the $k_{2}$-th block with size $\Delta_{2}$ and its $k_{1}$-th
sub-block, where the blocks are indexed by the following convention: If
$k_{l}=0$, then the block is on the main diagonal, otherwise is one of
the blocks off diagonal. In both cases, it is not important which block
is taken, as the circulant structure of the matrix leads to the same
result. For example, the element $B(1,0)$ of matrix~\eqref{P22}
corresponds to the off-diagonal element ($k_{1}=1$) of a diagonal block
($k_{2}=0$). Thus, $B(1,0)=1$.

The decomposition~\eqref{Jdec} proves that circulant matrices
are diagonalized by Transform~\eqref{F}, since the term inside the
parentheses is a diagonal matrix. Thus, Hamming matrices are
diagonalized by the same transform.

To arrive at formula~\eqref{spJ}, we need to show that
$B(k_{1},k_{2}) = f(2 - \delta_{k_1,0} - \delta_{k_2,0})$, which is
true for Hamming matrices but not for a general circulant matrix. We
begin with the observation that matrix $\mathcal
P_{(\Delta_{1},\ldots,\Delta_{L})}$ possesses a simple block structure:
The blocks on the diagonal are given by $\mathcal
P_{(\Delta_{1},\ldots,\Delta_{L-1})}$, whereas the blocks off diagonal
are given by the diagonal block but with all elements incremented by
one. For instance, for the case $\Delta=(2,2)$, we have that \be{}
    \mathcal P_{(2,2)} =
    \begin{bmatrix}
    \mathcal P_{(2)} & \mathcal P_{(2)} + \mathbf 1 \\
    \mathcal P_{(2)} + \mathbf 1 & \mathcal P_{(2)},
    \end{bmatrix}
\ee
where
\be{}
    \mathcal P_{(2)} =
    \begin{bmatrix}
    0 & 1 \\
    1 & 0
    \end{bmatrix}, \quad
\mathbf 1 =
    \begin{bmatrix}
    1 & 1 \\
    1 & 1
    \end{bmatrix}.
\ee

We can exploit the block structure of $\mathcal P_{\Delta}$
to obtain the form of the corresponding function
$B(k_{1},\ldots,k_{L})$. This function returns the element of matrix
$\mathcal P_{\Delta}$, which is an integer equal to the number of off
diagonal blocks necessary to locate the element. With our convention,
this is equal to the number of non-zero $k_{l}$s. For example, in the
$\mathcal P_{(\Delta_{1},\Delta_{2})}$ case, we have that
$B(k_{1},k_{2}) = 2 - \delta_{k_1,0} - \delta_{k_2,0}$.

For a general Hamming matrix, we can follow the above
reasoning, but replacing $d(I,J)$ with $f(d(I,J))$ for the element of
the matrix. As a consequence, the form of function $B(k_{1},k_{2})$ for
a general Hamming matrix of type $(\Delta_{1},\Delta_{2})$ reads
$B(k_{1},k_{2}) = f(2 - \delta_{k_1,0} - \delta_{k_2,0})$.
Formula~\eqref{spJ} in the main text is its generalization to a
sequences of arbitrary length $L$.

\section{Degeneracy structure of the spectrum of Hamming matrices}\label{B}
The aim of this Appendix is to prove that the leading eigenvalue of a Hamming matrix is non-degenerate if and only if the alphabet of the sequences is binary (i.e. $\Delta_{l}=2$, for each $l$). The proof is, as it stands, not useful for analyzing data. However, it enables us to prove a theorem that can be used in conjunction with the results of Sec.~\ref{sp_sec} and~\ref{exp_comp} to identify ecological drivers of niche diversification.

Let $\mathcal J$ be a Hamming matrix generated by a function $f$ as defined
in~\eqref{f}.  For each binary sequence ${\bf s} = (s_1,\dots,s_L)$,
where we write $|{\bf s}| = \sum_\ell s_\ell$, define the polynomial
  \begin{equation}\label{eq:defofp}
    p_{\bf s}(\alpha,\beta) = (\beta-\alpha)^{L-|{\mathbf s}|}
    \prod_{\ell = 1}^L ((\Delta_\ell - 1)\alpha + \beta)^{s_\ell}.
  \end{equation}
  Now, define the vector $\eta_{\bf s}$ so that $p_{\bf s}$ is its
  generating function, i.e.
\begin{equation}
  \label{eq:generating}
  p_{\bf s}(\alpha,\beta) = \sum_{k=0}^L \eta_{\bf s}^{(k)} \alpha^k
  \beta^{L-k}.
\end{equation}
Said another way, $\eta_{\bf s}^{(k)}$ is the coefficient of $\alpha^k$
in the polynomial $p_{\bf s}$.  Now define
\begin{equation}\label{eq:defoflambda}
  \lambda_{\bf s} = \sum_{k=0}^L f(k) \eta_{\bf s}^{(k)}.
\end{equation}
Then $\lambda_{\bf s}$ is an eigenvalue of $\mathcal J$ with multiplicity
\begin{equation}\label{eq:defofmu}
  \mu_{\bf s} = \prod_{l=1}^L (\Delta_{l}-1)^{1-s_{l}}.
\end{equation}

We prove these formulas below, but for now note that it follows from
this that if $\Delta_\ell > 2$ for all $\ell$, then the only non-degenerate
eigenvalue of $\mathcal J$ is $\lambda_{\bf 1}$.  We also show below that if $\mathcal J
= \mathcal{G}/C$ as defined in~\eqref{eq:defofC} and~\eqref{xlin},
then $\lambda_{\bf 1} = -1$.  In particular, this means that if
$\Delta_l >2$ for all $l$, then it follows that the unstable
eigenvalue is never non-degenerate, and the system is always in the degenerate
case described above.

Also, if $\Delta_l = \Delta_{l'}$ for some $l,l'$, and ${\bf s}, {\bf
  s'}$ are two vectors related by a transposition of the $l$-th and
$l'$-th coordinates, then $\lambda_{\bf s'} = \lambda_{\bf s}$.  In
particular, if all of the $\Delta_l$ are the same (call these numbers
$\Delta$), then $p_{\bf s}$ (and thus $\lambda_{\bf s}$) depends only
on $|{\bf s}|$, giving even more repeats.  For example, if $\Delta_l =
2$ for all $l$, then there are $L+1$ distinct eigenvalues, with
multiplicities given by $L!/k!(L-k)!$.  This is because even though
$\mu_{\bf s} = 1$ for all ${\bf s}$, $p_{\bf s}$ is the same for all
${\bf s}$ with the same number of ones.  For example, if $L=3$ and
$\Delta_l=2$ for all $l$,
\begin{equation}\label{eq:32}
\begin{split}
  \lambda_{111} &= f(0) + 3 f(1) + 3f(2) + f(3),\\
  \lambda_{110} = \lambda_{101} = \lambda_{011} &= f(0) + f(1) - f(2) - f(3),\\
  \lambda_{010} = \lambda_{100} = \lambda_{001} &= f(0)-f(1)-f(2)+f(3),\\
  \lambda_{000} &=  f(0) - 3 f(1) + 3f(2) - f(3).
\end{split}
\end{equation}
From this, we see that there are two ways in which we can obtain
multiple eigenvalues: We could have $\mu_{\bf s} > 1$ for some ${\bf
  s}$, or we could have $\lambda_{\bf s} = \lambda_{\bf s'}$ for two
different ${\bf s}, {\bf s'}$.

We will establish~\eqref{eq:defoflambda},~\eqref{eq:defofmu} in the case $L=2$, then discuss how the argument differs for larger $L$.
According to~\eqref{spJ}, when $L=2$ the eigenvalues are the numbers
$$
\sum_{k_1=0}^{\Delta_1-1} \sum_{k_2=0}^{\Delta_2-1} B(k_1,k_2) (\Omega_1^{k_1} \otimes \Omega_2^{k_2}),
$$
with $\Omega_l  = (1,\omega_l,\omega_l^2,\dots,
\omega_l^{\Delta_l-1})$.  Said another way, for any $\ell_1 = 0,\dots,
\Delta_1 -1$ and $\ell_2 = 0,\dots,\Delta_2 - 1$, the number
$$
\lambda_{\ell_1,\ell_2} = \sum_{k_1=0}^{\Delta_1-1} \sum_{k_2=0}^{\Delta_2-1} B(k_1,k_2) \omega_1^{\ell_1 \cdot k_1} \omega_2^{\ell_2\cdot k_2}
$$
is an eigenvalue.  For $L=2$,
$$
B(k_1,k_2) = \begin{cases} 0, & k_1=k_2=0,\\ 1 & k_1=0,k_2\neq 0\vee k_1\neq0,k_2=0,\\ 2,&k_1\neq0,k_2\neq0.\end{cases}
$$
This means that
\begin{equation*}
\begin{split}
\lambda_{\ell_1,\ell_2}
&= f(0) + f(1) \left(\sum_{k_1=1}^{\Delta_1-1} \omega_1^{\ell_1 \cdot k_1}+ \sum_{k_2=1}^{\Delta_2-1} \omega_2^{\ell_2 \cdot k_2}\right) \\
&\quad + f(2) \left(\sum_{k_1=1}^{\Delta_1-1} \sum_{k_2=1}^{\Delta_2-1}\omega_1^{\ell_1 \cdot k_1} \omega_2^{\ell_2\cdot k_2}\right).
\end{split}
\end{equation*}
We need the following identity:
$$
\sum_{k=1}^{\Delta-1} (\exp(2\pi i/\Delta))^{k \ell} = \begin{cases} \Delta-1, & \ell = 0\mod \Delta,\\ -1,& \mbox{else.}\end{cases}
$$ From this, we see that there are four cases for the formula for
$\lambda_{\ell_1,\ell_2}$, depending on whether or not $\ell_1,\ell_2$
are zero or nonzero:

\begin{align*}
f(0) &+ f(1) (\Delta_1+ \Delta_2 - 2) + f(2) (\Delta_1-1)(\Delta_2-1), \\
&\quad (\ell_1 = \ell_2 = 0),\\
f(0) &+ f(1) (\Delta_2 - 1) + f(2) (1-\Delta_1), \quad (\ell_1 \neq 0, \ell_2 = 0),\\
f(0) &+ f(1) (\Delta_1 - 1) + f(2) (1-\Delta_2), \quad (\ell_1 = 0, \ell_2 \neq 0),\\
f(0) &+ f(1) (-2) + f(2) , \quad (\ell_1, \ell_2 \neq 0),
\end{align*}
and we see that this matches the coefficients of the four polynomials:
\begin{equation}
\begin{split}
&(\alpha + (\Delta_1-1)\beta)(\alpha + (\Delta_2-1)\beta),\\& (\alpha + (\Delta_1-1)\beta)(\alpha-\beta), \\& (\alpha-\beta)(\alpha + (\Delta_2-1)\beta),\\&(\alpha-\beta)^2,
\end{split}
\end{equation}
respectively.  Also, we can see from counting that the four different
numbers have multiplicities $1$, $\Delta_1-1$, $\Delta_2-1$, and
$(\Delta_1-1)(\Delta_2-1)$, respectively.

Finally, we claim that if the matrix is defined as in~\eqref{xlin},
then the eigenvalue $\lambda_{\bf 1} = -1$.  To see this, notice that
the $\ell_1=\ell_2 = 0$ term above is
\begin{equation*}
f(0) + f(1) (\Delta_1+ \Delta_2 - 2) + f(2) (\Delta_1-1)(\Delta_2-1).
\end{equation*}
Notice that this must be the carrying capacity $C$, since each row of
$G$ must have one term of size $f(0)$, $\Delta_1+\Delta_2 - 1$ terms
of size $f(1)$ (this is the number of sequences that are unit Hamming
distance from any given sequence), and $(\Delta_1-1)(\Delta_2-1)$
terms of size $f(2)$ (similarly, the number of sequences distance two
from any given sequence).  Therefore the largest eigenvalue of
$\mathcal G/C$ is $1$ and thus the smallest eigenvalue of $-\mathcal
G/C$ is $-1$.  Moreover, if $\Delta_1,\Delta_2>2$, then it is clear
from the multiplicities that any positive eigenvalue of
$-\mathcal{G}/C$ is a multiple eigenvalue, giving rise to a degenerate
instability.

\end{appendix}
\end{document}